\documentclass[12pt]{article}%
\usepackage{amsmath}
\usepackage{cite}
\usepackage{graphicx}
\usepackage{amsfonts}
\usepackage{amssymb}
\usepackage{appendix}%
\providecommand{\U}[1]{\protect\rule{.1in}{.1in}}

\csname @addtoreset\endcsname{equation}{section}
\newcommand{\eq}{\begin{equation}}
\newcommand{\feq}{\end{equation}}
\newcommand{\eqn}{\begin{eqnarray}}
\newcommand{\feqn}{\end{eqnarray}}
\newcommand{\arr}{\begin{eqnarray*}}
\newcommand{\farr}{\end{eqnarray*}}

\newcommand{\bea}{\begin{eqnarray}}
\newcommand{\eea}{\end{eqnarray}}

\begin{document}
\begin{titlepage}
\begin{center}
\renewcommand{\thefootnote}{\fnsymbol{footnote}}
{\Large{\bf Kasner Branes with Arbitrary Signature.}}
\vskip1cm
\vskip 1.3cm
W. A. Sabra
\vskip 1cm
{\small{\it
Department of Physics\\
American University of Beirut\\ Lebanon \\}}
\end{center}
\bigskip
\begin{center}
{\bf Abstract}
\end{center}
We present static and time-dependent solutions for the theory of gravity  with a dilaton field and an arbitrary rank antisymmetric tensor.
The solutions constructed are valid for arbitrary space-time dimensions and signatures.
\end{titlepage}

\section{Introduction}

Many years ago, Kasner \cite{kasner} presented Euclidean four-dimensional
vacuum solutions which depend only on one variable. Those solutions can be
written in the form
\begin{equation}
ds^{2}=x_{1}^{2a_{1}}dx_{1}^{2}+x_{1}^{2a_{2}}dx_{2}^{2}+x_{1}^{2a_{3}}%
dx_{3}^{2}+x_{1}^{2a_{4}}dx_{4}^{2}.
\end{equation}
where the constants appearing in the metric, the Kasner exponents, satisfy the
following conditions
\begin{align}
a_{2}+a_{3}+a_{4}  &  =1+a_{1},\nonumber\\
a_{2}^{2}+a_{3}^{2}+a_{4}^{2}  &  =\left(  1+a_{1}\right)  ^{2}.
\label{kascon}%
\end{align}
The Kasner metric is actually valid for all space-time signatures
\cite{harvey}. If one writes%

\begin{equation}
ds^{2}=\epsilon_{0}x_{1}^{2a_{1}}dx_{1}^{2}+\epsilon_{1}x_{1}^{2a_{2}}%
dx_{2}^{2}+\epsilon_{2}x_{1}^{2a_{3}}dx_{3}^{2}+\epsilon_{3}x_{1}^{2a_{4}%
}dx_{4}^{2}, \label{kastwo}%
\end{equation}
where $\epsilon_{i}$ takes the values $\pm1,$ then the metric (\ref{kastwo})
can be shown to be a vacuum solution for four-dimensional gravity with all
possible space-time signatures provided the Kasner conditions (\ref{kascon})
are satisfied.

Dynamical time-dependent cosmological solutions can be obtained by simply
setting $\epsilon_{0}=-1$ and $\epsilon_{1}=\epsilon_{2}=\epsilon_{3}=1,$ and
after relabelling of the coordinates we obtain%

\begin{equation}
ds^{2}=-t^{2a_{1}}dt^{2}+t^{2a_{2}}dx^{2}+t^{2a_{2}}dy^{2}+t^{2a_{3}}dz^{2}.
\label{kasold}%
\end{equation}
Static solutions can also be obtained and are given by%

\begin{equation}
ds^{2}=-x^{2a_{2}}dt^{2}+x^{2a_{1}}dx^{2}+x^{2a_{3}}dy^{2}+x^{2a_{4}}dz^{2}.
\end{equation}
After a redefinition of the coordinate $t$, the metric (\ref{kasold}) can take
the following form
\begin{equation}
ds^{2}=-d\tau^{2}+\tau^{2a}dx^{2}+\tau^{2b}dy^{2}+\tau^{2c}dz^{2} \label{fk}%
\end{equation}
with%

\begin{equation}
a+b+c=a^{2}+b^{2}+c^{2}=1.
\end{equation}
This metric is what is normally referred to in the literature as the Kasner
metric. The Kasner metric is closely related to solutions found earlier by
Weyl \cite{weyl}, Levi-Civita \cite{levi} and Wilson \cite{wilson} and was
subsequently rediscovered by many authors \cite{harvey2}. We note that the
Kasner metric with two vanishing exponents corresponds to flat space-time,
where $t=0$ is only a coordinate singularity. Any other solution must have two
of the exponents positive and the third negative. As such, the Kasner metric
describes a homogenous universe which is expanding in two directions and
contracting in the third. In terms of Bianchi classification of homogenous
spaces, the Kasner metric corresponds to Bianchi type I. Generalised Kasner
metrics played a central part in the study of the cosmological singularity,
gravitational turbulence and chaos (see \cite{book} and references therein).

The Kasner vacuum solution can be generalised to arbitrary space-time
dimensions and signatures. In $d$ dimensions, we have
\begin{equation}
ds^{2}=\epsilon_{0}d\tau^{2}+%
{\displaystyle\sum\limits_{i=1}^{d-1}}
\epsilon_{i}\tau^{2a_{i}}dx_{i}^{2} \label{kas}%
\end{equation}
with the conditions
\begin{equation}%
{\displaystyle\sum\limits_{i=1}^{d-1}}
a_{i}=%
{\displaystyle\sum\limits_{i=1}^{d-1}}
a_{i}^{2}=1. \label{kgcon}%
\end{equation}
Let us now consider a $d$-dimensional gravity theory with an $m$-form $F_{m},$
with the action
\begin{equation}
S=\int d^{d}x\sqrt{\left\vert g\right\vert }\left(  R-\frac{\epsilon}%
{2\,m!}\,\,F_{m}^{2}\right)  . \label{gact}%
\end{equation}
where we allow for the possibility of a non-canonical sign of the kinetic term
of the $m$-form, $\epsilon=\pm1.$ The equations of motion derived from the
action (\ref{gact}) are given by
\begin{align}
R_{\mu\nu}-\epsilon\left(  \frac{1}{2(m-1)!}F_{\mu\alpha_{2}\cdots\alpha_{m}%
}F_{\nu}{}^{\alpha_{2}\cdots\alpha_{m}}-g_{\mu\nu}\frac{\left(  m-1\right)
}{2m!(d-2)}F_{m}^{2}\,\right)   &  =0,\nonumber\\
\partial_{\mu}\left(  \sqrt{\left\vert g\right\vert }\,\,F^{\mu\nu_{2}%
\cdots\nu_{m}}\right)   &  =0. \label{geq}%
\end{align}
For the metric of (\ref{kas}), the nonvanishing component of the Ricci tensor
are given by%
\begin{align}
R_{\tau\tau}  &  =\frac{1}{\tau^{2}}\sum_{i=1}^{d-1}\left(  a_{i}-a_{i}%
^{2}\right)  ,\nonumber\\
R_{x_{i}x_{i}}  &  =\epsilon_{0}\epsilon_{i}\tau^{2a_{i}-2}a_{i}\left(
1-\sum_{i=1}^{d-1}a_{i}\right)  .
\end{align}
If we consider solutions with the metric (\ref{kas}) and with the $m$-form
\begin{equation}
F_{m}=Pdx_{1}\wedge dx_{2}\wedge...\wedge dx_{m},
\end{equation}
then the equations of motion are satisfied provided%

\begin{align}
P^{2}  &  =\epsilon\epsilon_{0}\epsilon_{1}...\epsilon_{m}\left(
\frac{2\left(  m-1\right)  (d-2)}{m^{2}(d-m-1)}\right)  ,\nonumber\\
a_{1}  &  =a_{2}=..=a_{m}=\frac{1}{m},\nonumber\\
a_{m+1}  &  =...=a_{d-1}=-\frac{m-1}{m(d-m-1)}.
\end{align}
In four-dimensions with a Maxwell field, we simply get%

\begin{align}
P^{2}  &  =\epsilon\epsilon_{0}\epsilon_{1}\epsilon_{2}\nonumber\\
a_{1}  &  =a_{2}=\frac{1}{2},\text{ }a_{3}=-\frac{1}{2}. \label{ex}%
\end{align}
For dynamical solutions with Lorentzian signature where $-\epsilon
_{0}=\epsilon_{1}=\epsilon_{2}=\epsilon_{3}=1$, one must have $\epsilon=-1$,
corresponding to the theory with the non-canonical sign of the Maxwell term in
the action. The resulting solution obtained admits Killing spinors when the
theory is viewed as the bosonic sector of $N=2$ supergravity \cite{phantom}.
In four dimensions we can also have dyonic solutions, with the two-form given
by
\begin{equation}
F=Q\tau^{a_{3}-a_{1}-a_{2}}d\tau\wedge dz+Pdx\wedge dy
\end{equation}
where $P$ and $Q$ are constant. Then for the metric with exponents given in
(\ref{ex}), we obtain the condition
\begin{equation}
\epsilon\left(  \epsilon_{0}\epsilon_{1}\epsilon_{2}P^{2}-Q^{2}\epsilon
_{3}\right)  =1.
\end{equation}
If we consider $d$-dimensional gravity with a dynamical scalar field
\begin{equation}
S=\int d^{d}x\sqrt{\left\vert g\right\vert }\left(  R-\frac{1}{2}\partial
_{\mu}\phi\partial^{\mu}\phi\right)
\end{equation}
Then the metric (\ref{kas}) is a solution provided
\begin{align}
\phi &  =d_{1}\log\tau+d_{2},\\
\sum_{i=1}^{d-1}a_{i}  &  =\sum_{i=1}^{d-1}a_{i}^{2}+\frac{1}{2}d_{1}^{2}=1
\end{align}
where $d_{1}$ and $d_{2}$ are constants.

Static D-brane as well as time-dependent solutions (see for example \cite{gal}
and references therein) in M and string theories are of importance for the
study of duality symmetries, (A)dS/conformal field theory correspondence,
stringy cosmological models and cosmological singularities. Our aim in this
work is to find general Kasner-like brane solutions, which include static and
time-dependent solutions, in gravitational theories with an arbitrary rank
antisymmetric tensor field and a scalar dilaton. The general solutions can
then be used to construct explicit solutions for all supergravity theories
with non-trivial form fields and dilaton with various space-time signatures

\section{Brane Solutions}

We start with $d$-dimensional gravity theory with an $m$-form $F_{m}$ with the
action given in (\ref{gact}). We consider the following generic metric
solution
\begin{equation}
ds^{2}=e^{2U\left(  \tau\right)  }\left(  \epsilon_{0}d\tau^{2}+%
{\displaystyle\sum\limits_{i=1}^{p}}
\epsilon_{i}\tau^{2a_{i}}dx_{i}^{2}\right)  +e^{2V\left(  \tau\right)
}\left(
{\displaystyle\sum\limits_{j=p+1}^{d-1}}
\epsilon_{j}\tau^{2a_{j}}dx_{j}^{2}\right)  , \label{gm}%
\end{equation}
where $\epsilon_{0}$, $\epsilon_{i},$ $\epsilon_{j}$ take the values $\pm1$
and $a_{i}$, $a_{j}$ are all constants. Clearly we have $d=p+q+1.$ As in the
Kasner vacuum solutions with arbitrary space-time signature, dynamical
cosmological solutions as well as static solutions can be obtained depending
on whether $\tau$ is considered as a time or a spatial coordinate. Our metric
solution (\ref{gm}) is, to a great extent, motivated by the results of
\cite{kt}. There a correspondence between Melvin magnetic fluxtubes
\cite{melvin} and anisotropic cosmological solutions in four-dimensions was
explored. Our static solutions can therefore be thought of as generalizations
of Melvin fluxtubes to higher dimensions and with antisymmetric tensors and
dilaton field switched on.

\bigskip The non-vanishing components of the Ricci tensor for the metric
(\ref{gm}) are given by%

\begin{align}
R_{\tau\tau}  &  =-q\ddot{V}-p\ddot{U}-q\dot{V}\left(  \dot{V}-\dot{U}\right)
-\frac{1}{\tau}\left(  \left(  s-l\right)  \dot{U}+2l\dot{V}\right)  -\frac
{1}{\tau^{2}}\sum_{k=1}^{d-1}\left(  a_{k}^{2}-a_{k}\right)  ,\nonumber\\
R_{x_{i}x_{i}}  &  =-\epsilon_{0}\epsilon_{i}\mathrm{\tau}^{2a_{i}}\left[
\ddot{U}-\frac{a_{i}}{\tau^{2}}+\left(  \dot{U}+\frac{a_{i}}{\tau}\right)
\left(  (p-1)\dot{U}+q\dot{V}+\frac{l+s}{\tau}\right)  \right]  ,\nonumber\\
R_{\text{ }x_{j}x_{j}}  &  =-\epsilon_{0}\epsilon_{j}e^{2V-2U}\tau^{2a_{j}%
}\left[  \ddot{V}-\frac{a_{j}}{\tau^{2}}+\left(  \dot{V}+\frac{a_{j}}{\tau
}\right)  \left(  q\dot{V}+\left(  p-1\right)  \dot{U}+\frac{l+s}{\tau
}\right)  \right]  ,
\end{align}
where we have defined
\begin{equation}
l=\sum_{j=p+1}^{d-1}a_{j},\text{ \ \ }s=\sum_{i=1}^{p}a_{i}.
\end{equation}
It can be easily seen that a major simplification significant occurs if in
addition to keeping the Kasner conditions%

\begin{equation}
\sum_{k=1}^{d-1}a_{k}=\sum_{k=1}^{d-1}a_{k}^{2}=1, \label{kc}%
\end{equation}
we also impose the relation%

\begin{equation}
qV+(p-1)U=0. \label{nr}%
\end{equation}
The Ricci tensor non-vanishing components then take the much simpler form%

\begin{align}
R_{\tau\tau}  &  =-\ddot{U}-\left[  1-2\left(  \frac{d-2}{q}\right)  l\right]
\frac{\dot{U}}{\tau}-\frac{(p-1)(d-2)}{q}\dot{U}^{2},\nonumber\\
R_{x_{i}x_{i}}  &  =-\epsilon_{0}\epsilon_{i}\tau^{2a_{i}}\left(  \ddot
{U}+\frac{\dot{U}}{\tau}\right)  ,\nonumber\\
R_{x_{j}x_{j}}  &  =\frac{\left(  p-1\right)  }{q}\epsilon_{0}\epsilon
_{j}e^{2(\frac{2-d}{q})U}\tau^{2a_{j}}\left(  \ddot{U}+\frac{\dot{U}}{\tau
}\right)  .
\end{align}
We shall now consider solutions with a $p$-form given by%

\begin{equation}
F_{p}=Pdx_{1}\wedge dx_{2}\wedge...\wedge dx_{p}%
\end{equation}
with constant $P$. Then the Einstein equations of motion (\ref{geq}) reduce to
the two equations%

\begin{align}
\ddot{U}+\left(  1-2l\right)  \frac{\dot{U}}{\tau}+(p-1)\dot{U}^{2}  &
=0,\label{q1}\\
\ddot{U}+\frac{\dot{U}}{\tau}+\epsilon\epsilon_{0}\epsilon_{1}...\epsilon
_{p}\frac{qe^{2(1-p)U}}{2\left(  d-2\right)  }P^{2}\mathrm{\tau}^{-2s}  &  =0.
\label{q2}%
\end{align}
A solution for the equation (\ref{q1}) is given by
\begin{equation}
e^{U}=\left(  c_{1}+c_{2}\tau^{2l}\right)  ^{\frac{1}{p-1}}%
\end{equation}
which upon substitution in (\ref{q2}) gives the condition%
\begin{equation}
\frac{8\left(  d-2\right)  l^{2}}{q\left(  p-1\right)  }c_{2}c_{1}%
+\epsilon\epsilon_{0}\epsilon_{1}...\epsilon_{p}P^{2}=0.
\end{equation}
One can also consider (the dual) solutions with a $q+1$-form given by%

\begin{equation}
F_{q+1}=Q\,e^{2(1-p)U}\tau^{2l-1}d\tau\wedge dx_{p+1}\wedge...\wedge dx_{d-1}.
\end{equation}
In this case, we get the same form of solution with the various constants satisfying%

\begin{equation}
\frac{8\left(  d-2\right)  l^{2}}{q\left(  p-1\right)  }\epsilon\epsilon
_{p+1}...\epsilon_{d-1}c_{2}c_{1}-Q^{2}\,=0.
\end{equation}
As examples we consider eleven-dimensional supergravity theories with $(t,s)$
signatures with $t$ being the number of the time directions and $s$ the
spatial ones. The relevant action \cite{cjs, Hull} is
\begin{equation}
S=\int d^{11}x\sqrt{\left\vert g\right\vert }\left(  R-\frac{\epsilon}%
{48}\,F_{4}^{2}\right)  +...
\end{equation}
where $\epsilon$ takes the value $1$ for the theories with space-time
signatures $\left(  1,10\right)  $ , $\left(  5,6\right)  $ and $\left(
9,2\right)  ,$ and $\epsilon=-1$ for the mirror theories with signatures
$\left(  10,1\right)  $, $\left(  6,5\right)  $ and $\left(  2,9\right)  $.
Using our general results, solutions for all space-time signatures in eleven
dimensional supergravity theories can be constructed. For instance one can
obtain the time-dependent solution for the standard $\left(  1,10\right)  $
theory ($\epsilon=1),$ given by%
\begin{align}
ds^{2}  &  =e^{2U}\left(  -d\tau^{2}+\tau^{2a_{1}}dx_{1}^{2}+\tau^{2a_{2}%
}dx_{2}^{2}+\tau^{2a_{3}}dx_{3}^{2}+\tau^{2a_{4}}dx_{4}^{2}\right)
+e^{-U}\left(
{\displaystyle\sum\limits_{j=5}^{10}}
\tau^{2a_{j}}dx_{j}^{2}\right) \nonumber\\
F_{4}  &  =Pdx_{1}\wedge dx_{2}\wedge dx_{3}\wedge dx_{4}%
\end{align}
with%

\begin{equation}
e^{U}=\left(  1+c_{2}\tau^{2\sum_{j=5}^{10}a_{j}}\right)  ^{\frac{1}{3}%
},\text{ \ }P^{2}=4\left(  \sum_{j=5}^{10}a_{j}\right)  ^{2}c_{2}.
\end{equation}
Using the generic solution, setting $\epsilon_{0}=1,$ $\epsilon_{1}=-1$ and
relabelling the coordinates $\tau=x$ and $x_{1}=t,$ then we can obtain a
static solutions given by%
\begin{align}
ds^{2}  &  =e^{2U}\left(  -x^{2a_{1}}dt^{2}+dx^{2}+x^{2a_{2}}dx_{2}%
^{2}+x^{2a_{3}}dx_{3}^{2}+x^{2a_{4}}dx_{4}^{2}\right)  +e^{-U}\left(
{\displaystyle\sum\limits_{j=5}^{10}}
x^{2a_{j}}dx_{j}^{2}\right)  ,\nonumber\\
F_{4}  &  =Pdt\wedge dx_{2}\wedge dx_{3}\wedge dx_{4},
\end{align}
with
\begin{equation}
e^{U}=\left(  1+c_{2}x^{2\sum_{j=5}^{10}a_{j}}\right)  ^{\frac{1}{3}}%
,\ P^{2}=4\left(  \sum_{j=5}^{10}a_{j}\right)  ^{2}c_{2}%
\end{equation}
Similarly, we can also get the cosmological solutions%

\begin{align}
ds^{2}  &  =e^{2U}\left(  -d\tau^{2}+%
{\displaystyle\sum\limits_{i=1}^{7}}
\tau^{2a_{i}}dx_{i}^{2}\right)  +e^{-4U}\left(  \tau^{2a_{8}}dx_{8}^{2}%
+\tau^{2a_{9}}dx_{9}^{2}+\tau^{2a_{10}}dx_{10}^{2}\right)  ,\nonumber\\
F_{4}  &  =Q\,e^{-12U}\tau^{2\left(  a_{8}+a_{9}+a_{10}\right)  -1}d\tau\wedge
dx_{8}\wedge dx_{9}\wedge dx_{10},\text{ \ \ \ \ }Q^{2}\,=4l^{2}c_{2}\left(
a_{8}+a_{9}+a_{10}\right)  ^{2}\nonumber\\
e^{U}  &  =\left(  1+c_{2}\tau^{2\left(  a_{8}+a_{9}+a_{10}\right)  }\right)
^{\frac{1}{6}}.
\end{align}
For even-dimensional space-time one can consider "dyonic" solutions for
$p=q+1=\frac{d}{2}.$ The metric solutions for the $\frac{d}{2}$-form
\begin{equation}
F_{\frac{d}{2}}=Pdx_{1}\wedge dx_{2}\wedge...\wedge dx_{\frac{d}{2}%
}+Qe^{-U\left(  d-2\right)  }\tau^{2l-1}d\tau\wedge dx_{\frac{d}{2}+1}%
\wedge...\wedge dx_{d-1}%
\end{equation}
take the form%

\begin{equation}
ds^{2}=e^{2U}\left(  \epsilon_{0}d\tau^{2}+%
{\displaystyle\sum\limits_{i=1}^{\frac{d}{2}}}
\epsilon_{i}\mathrm{\tau}^{2a_{i}}dx_{i}^{2}\right)  +e^{-2U}\left(
{\displaystyle\sum\limits_{j=\frac{d}{2}+1}^{d-1}}
\epsilon_{j}\mathrm{\tau}^{2a_{j}}dx_{j}^{2}\right)  .
\end{equation}
The equations of motion in this case can be solved for
\begin{equation}
e^{U}=\left(  c_{1}+c_{2}\tau^{2l}\right)  ^{\frac{2}{d-2}}%
\end{equation}
where%
\begin{equation}
\epsilon\left(  Q^{2}\,\epsilon_{p+1}...\epsilon_{d-1}-P^{2}\epsilon
_{0}\epsilon_{1}...\epsilon_{p}\right)  =\frac{32l^{2}}{d-2}c_{1}c_{2}.
\end{equation}

As an example, we consider solutions of the standard Einstein-Maxwell theory
$\left(  \epsilon=1\right)  $, we get the cosmological solutions \cite{kt}%

\begin{align}
ds^{2}  &  =\left(  1+\frac{\left(  Q^{2}+P^{2}\right)  }{16a_{3}^{2}}%
\tau^{2a_{3}}\right)  ^{2}\left(  -d\tau^{2}+\tau^{2a_{1}}dx_{1}^{2}%
+\tau^{2a_{2}}dx_{2}^{2}\right)  +\left(  1+\frac{\left(  Q^{2}+P^{2}\right)
}{16a_{3}^{2}}\tau^{2a_{3}}\right)  ^{-2}\mathrm{\tau}^{2a_{3}}dx_{3}%
^{2}\nonumber\\
F_{2}  &  =Pdx_{1}\wedge dx_{2}+Qe^{-2U}\tau^{2a_{3}-1}d\tau\wedge dx_{3}%
\end{align}
as well as the static solutions%
\begin{align}
ds^{2}  &  =\left(  1+\frac{\left(  Q^{2}+P^{2}\right)  }{16a_{3}^{2}%
}x^{2a_{3}}\right)  ^{2}\left(  -x^{2a_{1}}dt^{2}+dx^{2}+x^{2a_{2}}dx_{2}%
^{2}\right)  +\left(  1+\frac{\left(  Q^{2}+P^{2}\right)  }{16a_{3}^{2}%
}x^{2a_{3}}\right)  ^{-2}x^{2a_{3}}dx_{3}^{2}\nonumber\\
F_{2}  &  =Pdt\wedge dx_{2}+Qe^{-2U}x^{2a_{3}-1}dx\wedge dx_{3}.
\end{align}

\bigskip

We now turn to the construction of solutions associated with an $m$-form
field, $F_{m}$, and a dilaton scalar, $\phi,$ coupled to the $m$-form field.
We take the action (in the Einstein frame)
\begin{equation}
S=\int d^{d}x\sqrt{\left\vert g\right\vert }\left(  R-\frac{1}{2}\partial
_{\mu}\phi\partial^{\mu}\phi-\frac{\epsilon}{2\,m!}\,e^{\beta\phi}\,F_{m}%
^{2}\right)  . \label{ga}%
\end{equation}
This represents the action of the bosonic fields of many supergravity
theories. We have allowed for the possibility of non-canonical sign of the
coupling term of the $F_{m}$ form ($\epsilon=\pm1)$ which will enable us to
construct solutions for various supergravity theories constructed by Hull
\cite{Hull}.

The equations of motion, derived from the variation of the action (\ref{ga})
are
\begin{align}
R_{\mu\nu}  &  =\frac{1}{2}\partial_{\mu}\phi\partial_{\nu}\phi+\frac{\epsilon
e^{\beta\phi}}{2(m-1)!}\left[  F_{\mu\alpha_{2}\cdots\alpha_{m}}F_{\nu}%
{}^{\alpha_{2}\cdots\alpha_{m}}-\frac{\left(  m-1\right)  }{m(d-2)}F_{m}%
^{2}\,g_{\mu\nu}\right]  ,\nonumber\\
\partial_{\mu}\left(  \sqrt{\left\vert g\right\vert }\,e^{\beta\phi}%
\,F^{\mu\nu_{2}\cdots\nu_{m}}\right)   &  =0,\nonumber\\
\frac{1}{\sqrt{\left\vert g\right\vert }}\,\partial_{\mu}\left(
\sqrt{\left\vert g\right\vert }\partial^{\mu}\phi\right)   &  =\frac{\beta
}{2\,m!}\epsilon e^{\beta\phi}F_{m}^{2}.
\end{align}
Let us consider solution with the $p$-form given by
\begin{equation}
F_{p}=Pdx_{1}\wedge dx_{2}\wedge...\wedge dx_{p}\text{ },
\end{equation}
where $P$ is a constant. We again take the metric (\ref{gm}) as our solution
supplemented with the constraints (\ref{kc}) and (\ref{nr}). The Einstein
equations of motion then give%

\begin{align}
\ddot{U}+\left[  1-2\left(  \frac{d-2}{q}\right)  l\right]  \frac{\dot{U}%
}{\tau}+\frac{(p-1)(d-2)}{q}\dot{U}^{2}  &  =-\frac{\dot{\phi}^{2}}{2}%
+\frac{\left(  p-1\right)  }{2(d-2)}j,\label{em1}\\
\ddot{U}+\frac{\dot{U}}{\tau}  &  =-\frac{q}{2\left(  d-2\right)
}j,\label{em2}\\
\,\frac{\dot{\phi}}{\tau}+\ddot{\phi}  &  =\frac{\beta}{2}j. \label{em3}%
\end{align}
where
\begin{equation}
j=\epsilon\epsilon_{0}\epsilon_{1}...\epsilon_{p}e^{2(1-p)U+\beta\phi}%
P^{2}\tau^{-2s}%
\end{equation}
The last two equations (\ref{em2}) and (\ref{em3}) imply that
\begin{equation}
\phi=-\frac{\beta(d-2)}{q}U. \label{su}%
\end{equation}
Upon substituting (\ref{su}) back in the equations (\ref{em1}) and (\ref{em2})
we obtain the following equations%

\begin{align}
\ddot{U}+\left(  1-2l\right)  \frac{\dot{U}}{\tau}+\mu\dot{U}^{2}  &
=0,\label{ba}\\
\ddot{U}+\frac{\dot{U}}{\tau}+\epsilon\epsilon_{0}\epsilon_{1}...\epsilon
_{p}\frac{qe^{-2\mu U}}{2\left(  d-2\right)  }P^{2}\tau^{-2s}  &  =0,
\label{bad}%
\end{align}
where we have defined $\mu=\left(  p-1+\frac{\beta^{2}}{2q}(d-2)\right)  .$
The equation (\ref{ba}) admits the solution
\begin{equation}
e^{U}=\left(  c_{1}+c_{2}\tau^{2l}\right)  ^{\frac{1}{\mu}} \label{metrics}%
\end{equation}
with the various constants satisfying the condition
\begin{equation}
\epsilon\epsilon_{0}\epsilon_{1}...\epsilon_{p}\frac{2(d-2)}{q}\frac{4l^{2}%
}{\mu}c_{2}c_{1}+P^{2}=0.
\end{equation}
We can also construct the dual solutions with
\begin{align}
F_{q+1}  &  =e^{-\beta\phi}Qe^{2(1-p)U}\tau^{2l-1}d\tau\wedge dx_{p+1}%
\wedge...\wedge dx_{d-1},\\
\phi &  =\beta\left(  \frac{d-2}{q}\right)  U.
\end{align}
and where
\begin{equation}
Q^{2}=\epsilon\epsilon_{p+1}...\epsilon_{d-1}\frac{8(d-2)}{q}\frac{l^{2}%
c_{1}c_{2}}{\mu}.
\end{equation}

Our analysis can be used to find non-trivial solutions for all the theories of
type IIA, type IIA$^{\ast}$, type IIB, \ type IIB$^{\ast}$ and type
IIB$^{\prime}$ supergravity theories with various space-time signatures
\cite{Hull}. In all those theories, the bosonic part consists of a dilaton and
a 3-form coming from the NS-NS sector. All type IIA and IIA$^{\ast}$ theories
also have a 2-form and a 4-form coming from the RR sector of the theory. The
type IIB, \ IIB$^{\ast}$ and IIB$^{\prime}$ have 1-form, 3-form and 5-form
coming from the RR sector. A list of these theories with corresponding
space-time signatures and signs for gauge kinetic terms can be found in
\cite{Hull}. For a given non-trivial RR field, the bosonic action in these
theories takes the form%

\begin{equation}
S=\int d^{10}x\sqrt{\left\vert g\right\vert }\left(  R-\frac{1}{2}%
\partial_{\mu}\phi\partial^{\mu}\phi-\frac{\epsilon}{2\,m!}\,e^{\frac{1}%
{2}\left(  5-m\right)  \phi}F_{m}^{2}\right)  .
\end{equation}
For a $p$-form given by%

\begin{equation}
F_{p}=Pdx_{1}\wedge dx_{2}\wedge...\wedge dx_{p}.
\end{equation}
A generic solution is given by%

\begin{equation}
ds^{2}=\left(  c_{1}+c_{2}\tau^{2l}\right)  ^{\frac{(9-p)}{8}}\left(
\epsilon_{0}d\tau^{2}+%
{\displaystyle\sum\limits_{i=1}^{p}}
\epsilon_{i}\tau^{2a_{i}}dx_{i}^{2}\right)  +\left(  c_{1}+c_{2}\tau
^{2l}\right)  ^{\frac{(1-p)}{8}}\left(
{\displaystyle\sum\limits_{j=p+1}^{9}}
\epsilon_{j}\tau^{2a_{j}}dx_{j}^{2}\right)
\end{equation}
with
\begin{equation}
\phi=-\frac{4\left(  5-p\right)  }{\left(  9-p\right)  }U.
\end{equation}
and with the various constants satisfying the condition%

\[
4\epsilon\epsilon_{0}\epsilon_{1}...\epsilon_{p}l^{2}c_{2}c_{1}+P^{2}=0.
\]
The dual solution is obtained with the $q+1$-form and scalar field given by%
\begin{align}
F_{q+1}  &  =Qe^{-\frac{32}{q}U}\tau^{2l-1}d\tau\wedge dx_{p+1}\wedge...\wedge
dx_{d-1},\\
\phi &  =4\left(  \frac{4}{q}-1\right)  U,\\
Q^{2}  &  =4\epsilon\epsilon_{p+1}...\epsilon_{d-1}l^{2}c_{1}c_{2}.
\end{align}

The self-dual $p=5$ case in various type IIB theories is treated separately.
In this case, we have $p=q+1=5$ and no dilaton coupling to $F_{5}$. Solutions
can be obtained using the general "dyonic" solutions constructed above. For
example, we take for the five-form%

\begin{equation}
F_{5}=P\left(  dx_{1}\wedge dx_{2}\wedge..\wedge dx_{5}+\epsilon_{1}%
\epsilon_{2}\epsilon_{3}\epsilon_{4}\epsilon_{5}e^{-8U}\tau^{2l-1}d\tau\wedge
dx_{6}\wedge dx_{7}\wedge dx_{8}\wedge dx_{9}\right)
\end{equation}
and the metric can take the form%

\begin{equation}
ds^{2}=e^{2U}\left(  \epsilon_{0}d\tau^{2}+%
{\displaystyle\sum\limits_{i=1}^{5}}
\epsilon_{i}\mathrm{\tau}^{2a_{i}}dx_{i}^{2}\right)  +e^{-2U}\left(
{\displaystyle\sum\limits_{j=6}^{9}}
\epsilon_{j}\mathrm{\tau}^{2a_{j}}dx_{j}^{2}\right)
\end{equation}
where%
\begin{align}
e^{2U}  &  =\left(  1+c_{2}\tau^{2\left(  a_{6}+a_{7}+a_{8}+a_{9}\right)
}\right)  ^{\frac{1}{2}}\label{d5}\\
\epsilon P^{2}\left(  \epsilon_{6}\epsilon_{7}\epsilon_{8}\epsilon
_{9}-\epsilon_{0}\epsilon_{1}\epsilon_{2}\epsilon_{3}\epsilon_{4}\epsilon
_{5}\right)   &  =4l^{2}c_{2}.
\end{align}

\bigskip

In this paper, we have constructed time-dependent and static solutions
generalising Kasner solutions to include matter fields. The solutions
constructed are for arbitrary space-time dimensions and signatures. The
formalism can be applied to find non-trivial cosmological and static solutions
in all known supergravity theories with form fields and a dilaton in various
space-time signatures. It is of importance to investigate the relevance of our
time-dependent solutions to string cosmology and the study of cosmological
singularities \cite{book}. We should also investigate the generalisation of
our results to find solutions in theories with many scalars and Maxwell
fields, such as $N=2$ supergravity theories in four and five dimensions. Work
along these lines is in progress and we hope to report on it in the near future.

\bigskip

\bigskip

{\flushleft{\textbf{Acknowledgements:}}}

\bigskip

This work is supported in part by the National Science Foundation under grant
number PHY-1620505.


\begin{thebibliography}{99}                                                                                               %


\bibitem {kasner}E. Kasner, \textit{Geometrical theorems on Einstein's
cosmological equations}, Am. J. Math. \textbf{43} (1921) 217.

\bibitem {harvey}A. Harvey, \textit{Complex Transformation of the Kasner
Metric, }General Relativity and Gravitation,  \textbf{21} (1989) 1021.

\bibitem {weyl}H. Weyl, \textit{H. Zur Gravitationstheorie}, Annalen der
Physik \textbf{54} (1917) 117.

\bibitem {levi}T. Levi-Civita, \textit{Rend. Acc. Lincei.} \textbf{26} (1917) 307.

\bibitem {wilson}W. Wilson, Phil. Mag. \textbf{40} (1928) 703.

\bibitem {harvey2}A. Harvey, \textit{Will the Real Kasner Metric Please Stand
Up}, General Relativity and Gravitation, \textbf{22}, $\left(  1990\right)
1433.$

\bibitem {book}V. Belinski, and Marc Henneaux, \textit{The Cosmological
singularity, }Cambridge University Press.

\bibitem {phantom}W. A. Sabra, \textit{Phantom Metrics With Killing Spinors},
Phys. Lett. \textbf{B750} (2015) 237; M. Bu Taam and W. A. Sabra,
\textit{Phantom space--times in fake supergravity, }Phys. Lett. \textbf{B751
}(2015) 297.

\bibitem {gal}C-M Chen, D. V. Gal'tsov and M. Gutperle, \textit{S brane
solutions in supergravity theories}, Phys. Rev. D \textbf{66} (2002) 024043.

\bibitem {kt}D. Kastor and J. Traschen, \textit{Melvin magnetic
fluxtube/cosmology correspondence, }Class. Quantum Grav. \textbf{32} (2015) 235027.

\bibitem {melvin}M. A. Melvin, \textit{Pure magnetic and electric geons,
}Phys. Lett. \textbf{8} (1964) 65.

\bibitem {cjs}E. Cremmer, B. Julia and J. Scherk, \textit{Supergravity Theory
in Eleven-Dimensions}, Phys.\ Lett.\ \textbf{B76} (1978) 409.

\bibitem {Hull}C. M. Hull, \textit{Duality and the signature of space-time,}
JHEP \textbf{11} (1998) 017.


\end{thebibliography}
\end{document}